\documentclass[aps,pra,reprint,footinbib,amsmath,amssymb,showpacs,floatfix,longbibliography]{revtex4-1}

\usepackage{graphicx,units,multirow,breakurl}
\usepackage{dcolumn}

\providecommand{\abs}[1]{\left\lvert#1\right\rvert}

\usepackage[colorlinks,allcolors=blue,hyperindex,breaklinks=true]{hyperref}

\begin{document}

\title{Searching for evidence of algorithmic randomness and incomputability\\in the output of quantum random number generators}
\author{John T.\ Kavulich}
\author{Brennan P.\ {Van Deren}}
\author{Maximilian Schlosshauer}
\address{Department of Physics, University of Portland, 5000 North Willamette Boulevard, Portland, Oregon 97203, USA}

\begin{abstract}
Ideal quantum random number generators (QRNGs) can produce algorithmically random and thus incomputable sequences, in contrast to pseudo-random number generators. However, the  verification of the presence of algorithmic randomness and incomputability is a nontrivial task. We present the results of a search for algorithmic randomness and incomputability in the output from two different QRNGs, performed by applying tests based on the Solovay--Strassen test of primality and the Chaitin--Schwartz theorem. The first QRNG uses measurements of quantum vacuum fluctuations. The second QRNG is based on polarization measurements on entangled single photons; for this generator, we use looped (and thus highly compressible) strings that also allow us to assess the ability of the tests to detect repeated bit patterns. Compared to a previous search for algorithmic randomness, our study increases statistical power by almost 3 orders of magnitude. \\[.2cm]
Journal reference: \emph{Phys.\ Lett.\ A} {\bf 388}, 127032 (2021),  DOI: \href{https://doi.org/10.1016/j.physleta.2020.127032}{\texttt{10.1016/j.physleta.2020.127032}}
\end{abstract}

\maketitle

\section{Introduction}

High-quality sources of random numbers are essential in a variety of areas, including cryptography \cite{Gisin:2002:ii}, simulations of complex systems  \cite{Motwani:1996:oo,Metropolis:1949:uu}, fundamental quantum experiments \cite{Brunner:2014:oo}, and information technology \cite{Hayes:2001:aa}. A typical approach for producing random numbers consists of feeding an initial seed value into a deterministic algorithm, which thus acts as a randomness expander. This approach is known as a pseudo-random number generator (PRNG). Sequences generated by PRNGs may be cyclic (i.e., have a finite period) or noncyclic, but by definition they will be (Turing) computable \cite{Turing:1937:oo,Calude:2010:kk}. The quality of the output from a PRNG is usually assessed by means of batteries of tests that look for the statistical distributions of predefined patterns in the output and then compare these distributions to those expected for an ideal random number generator. The NIST \cite{Rukhin:2010:ll} and Dieharder \cite{Brown:2020:za} test suites are widely used examples of such statistical random tests.

Tests of this kind, however, raise the question of how to define a pattern and how to make an exhaustive search for all possible patterns. Algorithmic information theory \cite{Chaitin:1977:km,Calude:2010:kk} provides a fundamental measure of randomness of (finite) strings and (infinite) sequences in terms of their Kolmogorov complexity \cite{Kolmogorov:1965:ii,Li:2008:ii}, which corresponds to the length of the shortest computer program that can reproduce the string or sequence. In this way, randomness is related to the incompressibility of strings and sequences \cite{Chaitin:1977:km,Calude:2010:kk}. This leads to the notion of algorithmic randomness \cite{Chaitin:1977:km,Calude:2010:kk} and to the definition of $c$-Kolmogorov random strings \cite{Calude:2010:kk}. Algorithmic randomness implies incomputatibility, but not vice versa. PRNGs will always produce computable sequences, which can therefore be expected to have low Kolmogorov complexity.

In contrast with a PRNG, which given the same seed will always produce the same sequence, a quantum random number generator (QRNG) \cite{Herrero:2017:kk} exploits the intrinsic physical randomness of quantum events to produce a string of bits. It is a basic tenet of quantum mechanics that even maximal knowledge of a quantum system does not allow one to predict the outcome of future measurements. This notion of irreducible ``quantum randomness''---the absence of predetermined ``values''---can be made precise, both theoretically and experimentally, in terms of violations of Bell-type inequalities for entangled states \cite{Brunner:2014:oo} and the value indefiniteness implied by the Kochen--Specker theorem \cite{Kochen:1967:hu}, leading to a \emph{certification} \cite{Pironio:2010:aa,Pironio:2013:qa,Fehr:2013:kn,Abbott:2012:kk,Abbott:2014:ll} of the random process. For the Bell-type approach, it was shown by Pironio \emph{et~al.} \cite{Pironio:2010:aa,Pironio:2013:qa} (see also Ref.~\cite{Fehr:2013:kn}) that, if we are given an initial amount of randomness and if we have a loophole-free realization of the experiment, the randomness and privacy of the output can be rigorously certified from Bell-inequality violations of the measurements used to produce the output. The Kochen--Specker theorem \cite{Kochen:1967:hu} demonstrates the impossibility of assigning definite values to all observables in systems of dimension $\ge 3$ in a noncontextual fashion \footnote{In fact, Refs.~\cite{Abbott:2014:hh,Abbott:2015:km} showed that only definite values of observables with probability-1 outcomes (as given by the Born rule) can be noncontextually assigned.}. The resulting value indefiniteness and indeterminism (which also give rise to a strong form of unpredictability \cite{Abbott:2015:qp}) were shown to provide randomness certification of QRNGs \cite{Calude:2008:za,Abbott:2012:kk,Abbott:2014:ll}, with an experimental realization described in Ref.~\cite{Kulikov:2017:uu}. In fact, it has been proven in Ref.~\cite{Abbott:2012:kk} that, with a mild additional physical assumption, infinite sequences produced from the outcomes of repeated measurements of a value-indefinite observable will \emph{with certainty} (that is to say, \emph{always}) be bi-immune \cite{Downey:2010:za}, a strong form of incomputability. It is not known \cite{Abbott:2015:qp}, however, whether there exists a QRNG that produces \emph{with certainty} (infinite) sequences that are also Martin-L\"of random \cite{MartinLof:1966:za,Calude:2010:kk}, a highly important form of algorithmic randomness \footnote{The emphasis in this and the preceding sentences is on the term ``with certainty.'' While any ideal random-number generator produces infinite sequences that are Martin-L{\"o}f random with probability one \cite{Calude:2010:kk}, it does not do so with certainty.}.

Instead of (or in addition to) implementing protocols that certify randomness and incomputability by demonstrating that the physical process by which the random output is produced is nonclassical, one may also look, \emph{a posteriori},  for evidence of algorithmic randomness (and thus incomputability) in the output generated by QRNGs. There are two difficulties: it is (Turing) undecidable whether a given string or sequence is ``truly'' random \cite{Li:2008:ii}, and there does not exist a single notion of randomness \cite{Calude:2010:kk}. One approach to this problem is to test the output of a QRNG for algorithmic randomness by assessing the performance of the random strings with respect to certain tasks (in the form of randomized algorithms), and then compare this performance to that of strings produced by PRNGs. In the context of QRNGs, such tests were first proposed and applied in Ref.~\cite{Calude:2010:oo}. Building on these preliminary investigations, in Ref.~\cite{Abbott:2019:uu} Abbott \emph{et~al.\ }developed four tests of algorithmic randomness that are based on the Solovay--Strassen test of primality \cite{Solovay:1977:yy} and the Chaitin--Schwartz theorem \cite{Chaitin:1978:um}; we will henceforth refer to them as Chaitin--Schwartz--Solovay--Strassen (CSSS) tests. The authors applied these tests to the output from a QRNG certified by Kochen--Specker value indefiniteness \cite{Kulikov:2017:uu}, as well as to the output from five PRNGs, using samples of 80 strings of $2^{26}$ bits each per generator. They performed a statistical comparison of the distributions of test results for the different RNGs, using the Kolmogorov--Smirnov test for two samples \cite{Hodges:1958:tv,Conover:1999:ii} and Welch's $t$-test \cite{Welch:1947:lm}. While they observed some instances of statistically significant differences between the QRNG on the one hand and the PRNGs on the other, the results were inconclusive and failed to demonstrate a clear advantage of the QRNG with respect to the algorithmic randomness and incomputability of its output.

Here we use the same set of four CSSS tests (adding to it a test of Borel normality) and statistical tests as in the study by Abbott \emph{et~al.\ }\cite{Abbott:2019:uu} just described, and apply them to strings from two other QRNGs, as well as to four PRNGs for comparison. While our methodology is therefore similar to that of  Ref.~\cite{Abbott:2019:uu}, we test substantially longer strings (by a factor of 25), and we also increase the size of the set of Carmichael numbers used in the CSSS tests by a factor of 33. This amplifies our statistical power by almost three orders of magnitude. For one of the QRNGs we test, we have available only comparably short strings, and we use looped versions of these strings. We turn this limitation into an asset: Because of their repetitive nature, these strings are highly compressible, and can therefore also be used as a tool for checking whether the randomness tests are capable of detecting such randomness-diminishing repetitions. 

In addition to the four CSSS tests, we apply a test of Borel normality to our six RNGs. While Borel normality cannot verify incomputability (there are Borel-normal but computable sequences \cite{Champernowne:1933:uu}), it is a necessary condition for algorithmic randomness and hence a useful property to check. Furthermore, QRNGs frequently exhibit bias effects that negatively affect the degree of Borel normality \cite{Calude:2010:oo,Martinez:2018:kk}, and hence this test is particularly pertinent in the context of QRNGs.

This paper is organized as follows. In Sec.~\ref{sec:methods}, we describe our RNGs, give a brief overview of the four CSSS tests and the test of Borel normality, and explain our methods for the statistical analysis of the test results. We present our results in Sec.~\ref{sec:results}, and offer a concluding discussion in Sec.~\ref{sec:discussion}.

\section{\label{sec:methods}Methods}

\subsection{Random number generators}

We used strings from two QRNGs, as well as strings from four PRNGs for comparison. To gather sufficient statistics, for each generator we took 100 strings containing $N=25 \times 2^{26} = 1,677,721,600$ bits each. For each of the four CSSS tests, in addition to the original strings we also tested the complemented strings in which all 0s and 1s are interchanged. The latter transformation of the string should leave properties of algorithmic randomness and incomputability intact \cite{Abbott:2019:uu}. Thus, if the test results differ in a statistically significant manner between two RNGs for the original (noncomplemented) strings but the difference disappears when the complemented strings are used (or vice versa), we have evidence that the observed difference in the output of the two RNGs should not be ascribed to properties of algorithmic randomness and incomputability. We will now briefly describe the RNGs used in this study.

\subsubsection{ANU QRNG}

The first QRNG, to be referred to as ``ANU QRNG'' in the following, is an experiment located at the Australian National University and based on measurements of quantum fluctuations of the vacuum. Technical details on the experiment can be found in Refs.~\cite{Symul:2011:cs,Haw:2015:im}. We used 200 files of random binary numbers that were generated in four consecutive runs and are publicly available from Ref.~\cite{ANU:2020:aa}. We combined pairs of consecutive files, resulting in a sample of 100 strings of $N=1,677,721,600$ bits each. The ANU QRNG is particularly useful in practice because of its very high data rate (\unit[5.7]{Gbits/s}) and the public accessibility of its output, including the availability of various software that directly interfaces with this output.  Furthermore, the statistical randomness of the output is confirmed through several continuously running batteries of tests, including the NIST \cite{Rukhin:2010:ll} and Dieharder \cite{Brown:2020:za} tests. We note that the randomness and privacy of the output from the ANU QRNG is not certified through loophole-free violations of the Bell--CHSH inequality \cite{Pironio:2010:aa,Pironio:2013:qa,Fehr:2013:kn} or the value indefiniteness implied by the Kochen--Specker theorem \cite{Kochen:1967:hu,Abbott:2012:kk,Calude:2008:za}. 

As part of the experimental procedure, the raw output is significantly postprocessed to extract nearly uniform random strings from the original Gaussian distribution with the aid of a cryptographic hashing algorithm \cite{Symul:2011:cs,Haw:2015:im}. Certain types of postprocessing can influence algorithmic randomness; for example, it has been shown that von Neumann postprocessing can diminish algorithmic randomness for (infinite) sequences \cite{Abbott:2012:xa}. Notwithstanding such potential effects, postprocessing is a necessary fact of most practically useful QRNGs. Therefore, if we want to check whether the output from such QRNGs exhibits an advantage over PRNGs in terms of algorithmic randomness, then the testing of such postprocessed output is an important concern, however such postprocessing may or may not influence properties of algorithmic randomness compared to the raw output.

\subsubsection{\label{sec:parity-qrng}Parity QRNG}

This QRNG is situated in our own lab and produces bits from joint measurements on polarization-entangled pairs of single photons created from type-I spontaneous parametric downconversion. For details on our experiment, we refer the reader to Ref.~\cite{Coleman:2020:za}. The random bits are generated in the course of a measurement of the Bell--CHSH inequality \cite{Clauser:1969:ii}. The measured violation is used to estimate (from the theorem of Pironio \emph{et~al.\ }\cite{Pironio:2010:aa}) a lower bound on the source min-entropy. Since we do not perform a loophole-free test of the Bell--CHSH inequality, the output from our QRNG cannot be considered rigorously certified in the sense of Refs.~\cite{Pironio:2010:aa,Pironio:2013:qa,Fehr:2013:kn}. Nevertheless, one may take the observed violations as an indication of the nonclassical nature of the measured photon statistics used to generate the random output, which is thus suggestive of the presence of quantum randomness. The output from our QRNG is not postprocessed in any way, and this raw output has previously been shown to pass all applicable statistical randomness tests in the NIST suite and to be Borel normal \cite{Coleman:2020:za}. 

Instead of time-tagging individual photon events, we count the numbers of photons registered at each of the four single-photon detectors during a given time interval $\tau=\unit[0.1]{s}$ and use the parity (even or odd) of these count numbers to produce four random bits per counting interval. Because of this approach, the bit rate in our experiment is low, on the order of 26~bits/s, and with our available time we were able to produce only a single string containing $6.4 \times 10^6$ bits. This is three orders of magnitude shorter than what we have available from the ANU QRNG (and the PRNGs). Therefore, we cannot hope to compare the output from our QRNG to the output from the ANU QRNG and the other PRNGs on the same footing. Yet, we chose to include it in our comparison to highlight how the results of the tests may or may not be sensitive to the issue of string length. In this sense, we wish to test not only the QRNG, but also the tests themselves. 

We break the string into 100 strings of $6.4 \times 10^4$ bits each. For the Borel test, we test these short strings directly (see Sec.~\ref{sec:borel-normality} for an explanation of this choice). For the CSSS tests, we repeatedly loop each string until it reaches the length $N=1,677,721,600$ bits of the strings from the other generators (this requires 26,215 looping steps, with the final $N$-bit string containing 26,214 complete repetitions of the original string). This means that each string contains only a comparably short unique string that is then repeated many times. Thus, the looped strings are highly compressible, and hence one would expect the looping to diminish both statistical and algorithmic randomness. Indeed, we find that these strings fail all but one of the Dieharder statistical tests \cite{Brown:2020:za}.

\subsubsection{PRNGs}
 
We used output from the PRNGs AES-OFB \cite{Dworkin:2001:ll}, Gfsr4 \cite{Ziff:1998:lm}, Mt19937 \cite{Matsumoto:1998:km}, and Threefish \cite{Ferguson:2010:oo}, all as implemented by Dieharder \cite{Brown:2020:za}. 

AES-OFB and Threefish are cryptographic block ciphers \cite{Dworkin:2001:ll,Ferguson:2010:oo}. Gfsr4 is a generalized feedback shift-register generator that behaves like a lagged-Fibonacci generator \cite{Ziff:1998:lm}. Mt19937 is a standard implementation of the Mersenne Twister based on the Mersenne prime $2^{19937}-1$ \cite{Matsumoto:1998:km}.

\subsection{Randomness tests}

We will now briefly describe the five randomness tests we have applied to our random strings: the test of Borel normality (Sec.~\ref{sec:borel-normality}) and the four different Chaitin--Schwartz--Solovay--Strassen tests (Sec.~\ref{sec:chait-schw-solov}).

\subsubsection{\label{sec:borel-normality}Borel normality}

Borel normality \cite{Calude:1994:hh,Calude:2010:kk} is a necessary condition for algorithmic randomness and incomputability. It is, however, not sufficient; perhaps the most well-known counterexample is Champernowne's constant \cite{Champernowne:1933:uu}, which is Borel normal but manifestly computable. Violations of Borel normality have previously been found in the outputs from QRNGs \cite{Calude:2010:oo,Martinez:2018:kk,Abbott:2019:uu}, and it has been suggested that this failure of Borel normality may have been caused by bias in the tested strings \cite{Martinez:2018:kk,Abbott:2019:uu}. 

Borel normality applied to a finite string evaluates whether all substrings of given length $m$ occur with the expected probability of $2^{-m}$ \cite{Calude:1994:hh,Calude:2010:kk}. It relates to the compressibility of the string by a lossless finite-state machine \cite{Ziv:1978:aa}. A string $\mathbf{x}$ of length $\abs{\mathbf{x}}$ is considered Borel normal if the following condition holds for all integer $m$ with $1 \le m \le \log_2 \log_2 \abs{\mathbf{x}}$ \cite{Calude:1994:hh,Calude:2010:kk}:
\begin{equation}\label{eq:borel}
\max_{1 \le j \le 2^m} \left| \frac{N_j^m(\mathbf{x})}{\abs{\mathbf{x}}/m} - \frac{1}{2^m} \right| \le \sqrt{\frac{\log_2 \abs{\mathbf{x}}}{\abs{\mathbf{x}}}},
\end{equation}
where $N_j^m(\mathbf{x})$ is the number of occurrences of the $j$th string drawn from the alphabet of all binary strings of length $m$. For our string lengths, the maximum value of $m$ is $m=4$. For the test metric, we rewrite Eq.~(\ref{eq:borel}) as
\begin{equation}\label{eq:borel2}
\left(\max_{1 \le j \le 2^m} \left| \frac{N_j^m(\mathbf{x})}{\abs{\mathbf{x}}/m} - \frac{1}{2^m} \right| \right) \sqrt{\frac{\abs{\mathbf{x}}}{\log_2 \abs{\mathbf{x}}}}\le 1,
\end{equation}
and record the value of the left-hand side for each string we test.

As noted in Sec.~\ref{sec:parity-qrng}, for the strings produced by the Parity QRNG, we use the unlooped 64,000-bits strings to evaluate Borel normality. This is so because the bound on the right-hand side of Eq.~(\ref{eq:borel}) depends on the string length, the idea being that the relative frequencies of the $m$-bit substrings will get closer to the expected value of $2^{-m}$ as the length of a given string increases. If we were to use the looped strings instead, then we would be imposing the tight bound associated with the full-length ($N=1,677,721,600$ bits) strings, while the relative frequencies of the $m$-bit substrings would still be those of the unlooped 64,000-bits strings, leading to a certain violation of Borel normality that is not indicative of the Borel normality of the output from the actual bit-generating process. (Indeed, we find that the looped strings violate Borel normality by a factor of $33 \pm 9$.)

\subsubsection{\label{sec:chait-schw-solov}Chaitin--Schwartz--Solovay--Strassen tests}

We apply four versions of the Chaitin--Schwartz--Solovay--Strassen (CSSS) tests as proposed and carried out in Ref.~\cite{Abbott:2019:uu}; we refer the reader to this reference for details. To implement the CSSS tests, we used computer code adapted from code made publicly available \cite{Abbott:sources} by the authors of Ref.~\cite{Abbott:2019:uu}. We will now briefly describe these tests.

\paragraph{Overview.} The CSSS tests use strings from RNG sources as ``witnesses'' to test whether or not a given integer $n$ is composite or (probably) prime. The first ingredient of the CSSS tests is the Solovay--Strassen test of primality \cite{Solovay:1977:yy}. For an integer $n$, the Solovay--Strassen test defines a predicate $W(n,a)$ given by 
\begin{equation}\label{eq:W}
\left(\frac{a}{n}\right) \not= a^{(n-1)/2}  \quad (\mathrm{mod}\, n),
\end{equation}
where $a$ is a natural number between 1 and $(n-1)$, and $\left(\frac{a}{n}\right)$ is the Jacobi symbol. $W(n,a)$ is called the Solovay--Strassen predicate. 

If $n$ is prime, then $W(n,a)$ is false for all $a \in [1, n-1]$. Thus, if $W(n,a)$ is true for some $a \in [1, n-1]$, then $n$ is composite, and we call $a$ a witness to the compositeness of $n$. If $W(n,a)$ is false for a given $a$, then the probability that $n$ is prime is larger than $\frac{1}{2}$; more generally, if we use $N$ such numbers $a_k$ uniformly distributed in $[1,n-1]$, and if $W(n,a_k)$ is false for all $1 \le k \le N$, then the probability that $n$ is prime is larger than $1-2^{-N}$. This is so because, as shown in Ref.~\cite{Solovay:1977:yy}, at least half of the numbers $a_k \in [1,n-1]$ are witnesses to the compositeness of $n$ [i.e., they will satisfy $W(n,a)$] if $n$ is composite, and none of these numbers will satisfy $W(n,a)$ if $n$ is prime. Thus, although the Solovay--Strassen test cannot confirm with certainty that a given integer $n$ is prime, the probability that $n$ is indeed prime increases rapidly with the size of the set of numbers $a_k$ that fail to witness the compositeness of $n$.

The second ingredient of the CSSS tests is the Chaitin--Schwartz theorem \cite{Chaitin:1978:um}. It transforms the probabilistic result of the Solovay--Strassen test into a rigorous proof of primality if the potential witnesses are derived from $c$-Kolmogorov random strings, in the following way. Let $s$ be a string of length $m$ in 2-bit binary representation, and let $n$ be an integer. Rewrite $s$ into base $(n-1)$, with digits $s=d_k d_{k-1} \cdots d_0$ over the alphabet $\{0,1,\dots,n-1\}$, where $k$ is given by the smallest integer that satisfies the inequality
\begin{equation}\label{eq:k}
k > \frac{\log(2^m-1)}{\log(n-1)}-1.
\end{equation}
Now define the compound predicate 
\begin{align}\label{eq:Z}
Z(n,s) &= \neg W(n,1+d_0) \wedge \neg W(n,1+d_1) \wedge  \cdots \notag \\  & \qquad \wedge \neg W(n,1+d_{k-1}),
\end{align}
where $W(n,1+d_i)$ is the Solovay--Strassen predicate~(\ref{eq:W}). Then the Chaitin--Schwartz theorem states that for all sufficiently large $c$, if $s$ is any $c$-Kolmogorov random string containing $\ell(\ell+2c)$ bits and $n$ is an integer whose binary representation is $\ell$ bits long, then $Z(n,s)$ is true if and only if $n$ is prime \cite{Chaitin:1978:um}. In other words, if we use the digits of $s$ (in base $n-1$) as a potential witness, then if the Solovay--Strassen predicates $W(n,1+d_i)$ formed from these individual digits are all false, then $n$ must be prime. 

Loosely speaking, we may thus say that $c$-Kolmogorov random strings perform, on average, better than nonrandom strings in testing primality, and thus in witnessing compositeness. In this way, the CSSS tests can serve to probe and detect the presence of algorithmic randomness and incomputability. The four CSSS tests of Ref.~\cite{Abbott:2019:uu} are based on these ideas. All of them take strings from the RNG source to form large sets of potential witnesses, and apply them to sets of composite numbers to check for their primality, making use of the above results. The tests then measure how well these strings perform in this primality-testing task, allowing us to compare the performance of strings from a given QRNG to the performance of strings from PRNGs. 

\paragraph{First and second CSSS tests} The first and second CSSS tests differ only in their choice of test metric. They directly use binary strings from the source as witnesses and look for how many witnesses (first test) or bits (second test) are required to verify the compositeness of a set of test numbers. As in Ref.~\cite{Abbott:2019:uu}, we choose Carmichael numbers as test numbers, using the set of all 8,220,777 Carmichael numbers up to $10^{20}$ (calculated in Ref.~\cite{Pinch:2007:km}). Carmichael numbers are odd composite numbers (having at least three prime factors), but they satisfy Fermat's little theorem and are hard to factorize, and therefore also referred to as pseudoprimes. For a given Carmichael number $n$, the tests read $\log_2(n)$ bits from the source. The corresponding integer in decimal representation forms the potential witness $a$ provided $a \le n-1$; if $a> n-1$, we discard $a$ and form a new potential witness by reading the next  $\log_2(n)$ bits from the source. Except for the strings from the Parity QRNG, our random strings are long enough to contain a number of witnesses sufficient for verifying the compositeness of all Carmichael numbers in the set without having to recycle bits. 

\paragraph{Third CSSS test} This test applies the procedure of the Chaitin--Schwartz theorem to form the witnesses. That is, for a given Carmichael number $n$ (again using all Carmichael numbers up to $10^{20}$), the test reads a string $s$ containing $m$ bits from the source, where $m=\ell(\ell+2c)$ with the choice $c=\ell-1$ (as in Ref.~\cite{Abbott:2019:uu}) and $\ell$ is the length of $n$ in 2-bit binary representation. Then $s$ is converted to base $(n-1)$, and the digits of $s$ in this representation form the witnesses as specified by Eq.~(\ref{eq:Z}). We then count how many bits of $s$ need to be used until one of the Solovay--Strassen predicates $W(n,1+d_i)$ in Eq.~(\ref{eq:Z}) is true, i.e., until $n$ is verified to be composite. In the unlikely event that all of the $W(n,1+d_i)$ are false and therefore $Z(n,s)$ is true, we have a situation that Abbott \emph{et~al.\ }\cite{Abbott:2019:uu} refer to as a ``violation of the  Chaitin--Schwartz theorem.'' It is a violation in the sense that if $s$ were a $c$-Kolmogorov random string (here with $c=\ell-1$), then $Z(n,s)$ being true would imply the primality of $n$, in contradiction with the fact that $n$ is a Carmichael number and hence composite. If such a violation occurs, then (as in Ref.~\cite{Abbott:2019:uu}) we count all the bits of $s$ for the purpose of the test metric. 

\paragraph{Fourth CSSS test} This test specifically looks for violations of the Chaitin--Schwartz theorem in the sense just defined. That is, for some arbitrary (not necessarily $c$-Kolmogorov random) string $s$, we look for instances in which the predicate $Z(s,n)$ is true [i.e., all the $W(n,1+d_i)$ are false]. Since for a $c$-Kolmogorov random string as specified by the Chaitin--Schwartz theorem $Z(s,n)$ will necessarily be false, such instances may be used to detect a lack of algorithmic randomness in the string $s$, or at least highlight a difference in performance between  $c$-Kolmogorov random strings and nonrandom strings in the test. 

As pointed out in Ref.~\cite{Abbott:2019:uu}, the problem with applying this search for violations in practice is that the probability of observing a violation is very low and decreases rapidly with the length of the number whose compositeness is tested. To see this, recall that for a given composite number $n$ to be tested, and $\ell$ denoting the length of $n$ in 2-bit binary representation, we read a string $s$ of length $m=\ell(3\ell-2)$ in 2-bit binary representation. The number of digits of $s$ in base $(n-1)$ is given by the smallest integer $k$ that fulfills the inequality (\ref{eq:k}). Then the Solovay--Strassen theorem implies that the probability for any such $s$ to produce a violation is smaller than $2^{-k}$. For example, for the smallest Carmichael number $n=561$, we have $\ell=10$ and $m=280$, and thus $k =30$, giving a probability less than $2^{-30} \approx 10^{-9}$. This means that, on average, we need to read more than $N_\mathrm{min} = 280 \times 2^{30} \approx 3 \times 10^{11}$ bits before we can hope to detect a single violation, which exceeds the length of our individual strings by a factor of 180.

Following the approach of Ref.~\cite{Abbott:2019:uu}, we address this issue in two ways. First, we focus on testing relatively small numbers $n$ only. We choose to test all odd composite numbers less than 100, and additionally include the smallest Carmichael number (561), giving the following set of 26 numbers:
\begin{eqnarray}\label{eq:testnos}
&&\{ 9, 15, 21, 25, 27, 33, 35, 39, 45, 49, 51, 55, 57, 63,\nonumber\\
&& \,\,\,65, 69, 75, 77, 81, 85, 87, 91, 93, 95, 99, 561 \}.
\end{eqnarray}
Second, we make multiple passes through each string with an incremental bit offset on each pass. Thus, on the first pass we start from the first bit, on the second pass from the second bit, and so on. For a given $\ell$-bit number $n$ to be tested and corresponding blocks of $m=\ell(3\ell-2)$ bits read from the source to form the witness strings, we make $m$ passes. This corresponds to numbers of passes between $m=40$ for $n=9$ and $m=280$ for $n=561$. After $m$ passes, we will have used up all the bits, since passes with larger offsets would not produce any fresh witness strings.

\subsection{Statistical analysis}

For each test, we look for statistically significant differences in the results for the different RNGs, by comparing the distributions of test results (henceforth referred to as ``datasets'') for each sample of 100 strings using the statistical analysis described in Ref.~\cite{Abbott:2019:uu}.  We first use the Kolmogorov--Smirnov test for two samples. This test is distribution-free, i.e., it neither makes nor needs assumptions about the nature of the underlying probability distribution. If the $p$-value produced by the test is larger than a chosen threshold, then we cannot reject the hypothesis that the distributions of the two samples are the same. As in Ref.~\cite{Abbott:2019:uu}, we choose a low threshold of $p=0.005$. Thus, we consider the difference between two datasets to be statistically significant if $p < 0.005$. For each of the four CSSS tests and for a given pair of RNGs, we compare the corresponding datasets first for the original strings, then for the complemented strings. Since we are particularly interested in discerning differences between the QRNGs and the PRNGs, we also compare the results for the complemented QRNG strings to the results for the original PRNG strings, and vice versa. 

In addition to the Kolmogorov--Smirnov test, we use Welch's $t$-test \cite{Welch:1947:lm} to look for statistically significant differences between the means of the datasets, using again the threshold $p=0.005$. Thus, we consider the difference in the means between two datasets to be statistically significant if $p<0.005$. Since this test presumes a normal distribution, before applying it we first use the Shapiro--Wilk test \cite{Shapiro:1965:ll}, which tests the null hypothesis that the data was drawn from a normal distribution. Again, the test produces a $p$-value, and we take evidence of non-normality to be $p < 0.05$. Only once normality has been assured for a pair of datasets do we proceed to apply the Shapiro--Wilk test.

\section{\label{sec:results}Results}

\subsection{Borel normality test}

\begin{figure}
\centering
\includegraphics[width=3.4in]{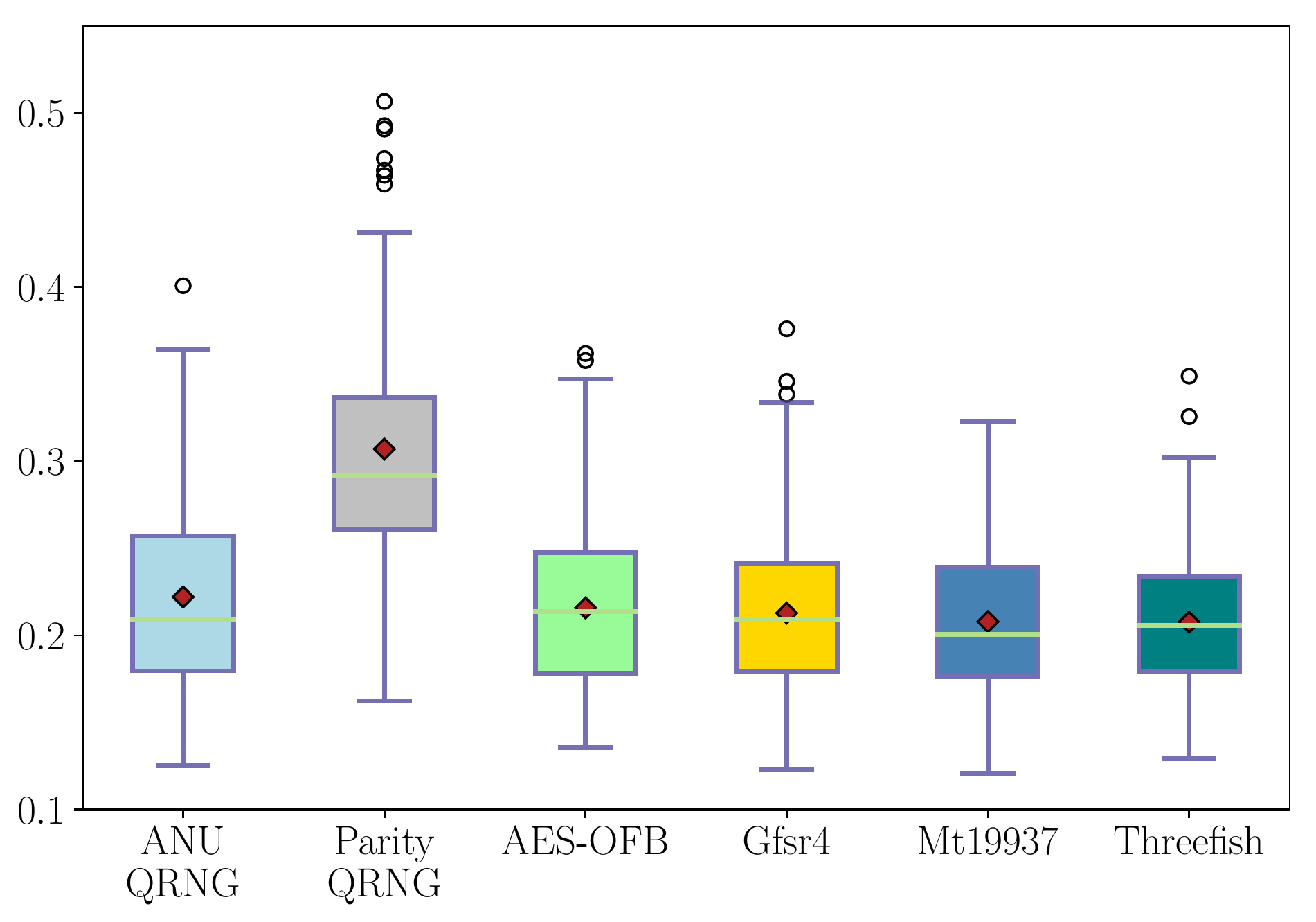} 
\caption{\label{fig:borel}Results of the Borel normality test. For the QRNG Parity generator, the set of 100 unique 16-kbit strings (without looping) was used for the test.}
\end{figure} 

The results from the Borel normality test for the two QRNGs and the four PRNGs are shown in Fig.~\ref{fig:borel}. There are two immediate observations. The first is that none of the strings violate Borel normality, since all test values lie well below 1. Second, it is evident that the distribution for the Parity QRNG differs significantly from the other RNGs. This, however, is unsurprising, since the strings from the Parity QRNG are orders of magnitude shorter than those from the other RNGs, and accordingly one expects stronger discrepancies between the expected frequencies of the $m$-bit ($m=1,2,3,4$) substrings and the observed frequencies. 

The Kolmogorov--Smirnov test readily confirmed the difference between the Parity QRNG dataset and the other five datasets, with $p$-values below $10^{-13}$. For all pairs of datasets for the other five generators, the Kolmogorov--Smirnov test failed to detect any statistically significant differences. The Shapiro--Wilk test indicated non-normality of all datasets except Threefish (which had a $p$-value of 0.053, just marginally above the threshold of 0.05). Thus, we did not apply Welch's $t$-test.

For $m=1$, testing Borel normality corresponds to checking for bias in the relative frequencies $p_0$ and $p_1$ of 0s and 1s in the string. For the ANU QRNG, we find a mean bias of $|p_0-0.5| = (9.8 \pm 0.8) \times 10^{-6}$, where the average is taken over the 100 strings in the sample. For the Parity QRNG, we find $|p_0-0.5| = 2.2 \times 10^{-5}$ for the string containing all $6.4 \times 10^6$ unique bits. Thus, in both cases the bias is very small. For the ANU QRNG, near-uniformity of the output arises from postprocessing. For the Parity QRNG, it is due to the use of the parity variable to generate the bits, which tends to wash out any experimentally introduced bias caused by factors such as imperfect state preparation, beam splitter crosstalk, and imbalanced detectors \cite{Herrero:2017:kk,Coleman:2020:za}.

\subsection{First Chaitin--Schwartz--Solovay--Strassen test}

\begin{figure}
\centering

{\small (a)} 

\includegraphics[width=3.4in]{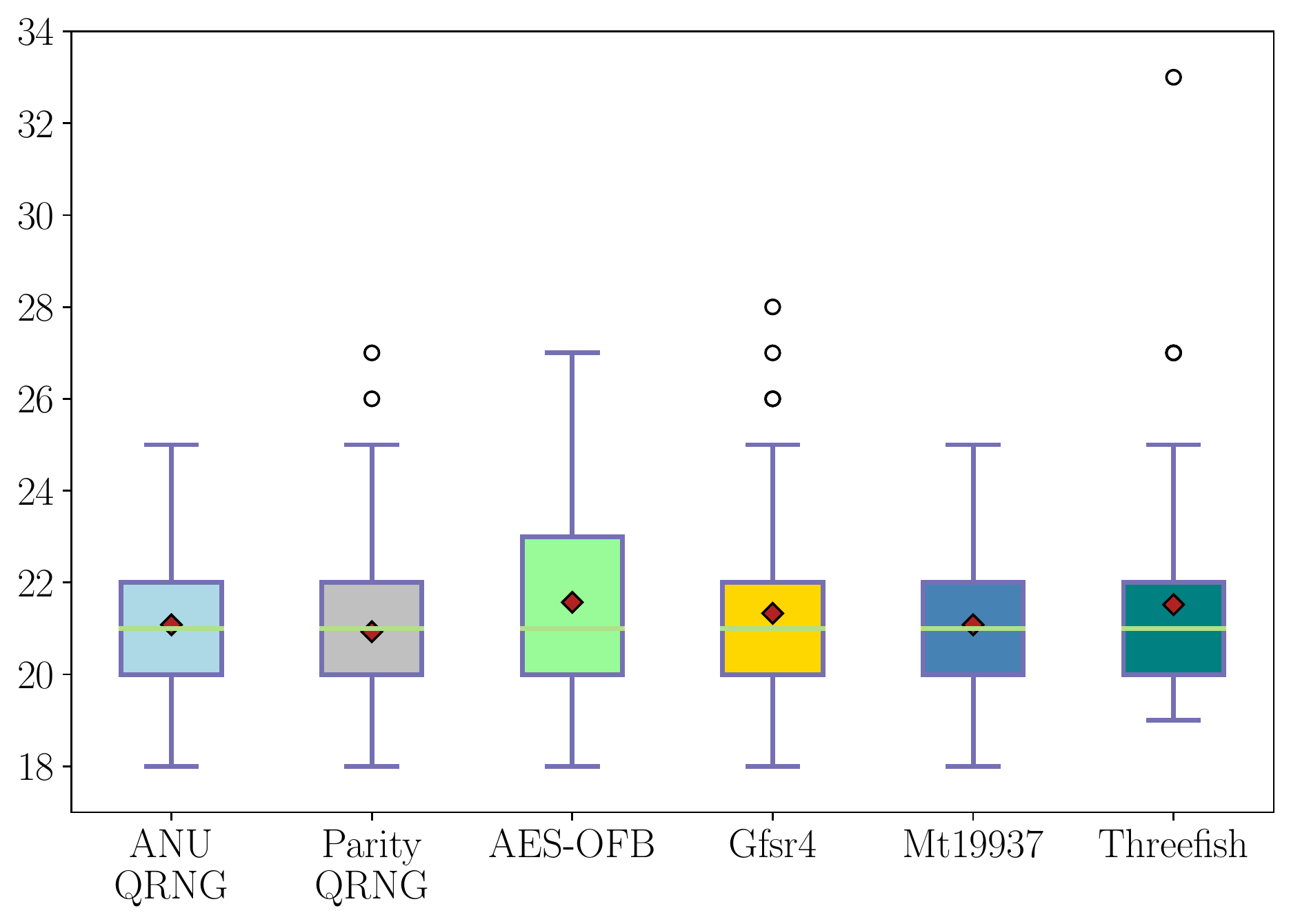}

{\small (b)}

 \includegraphics[width=3.4in]{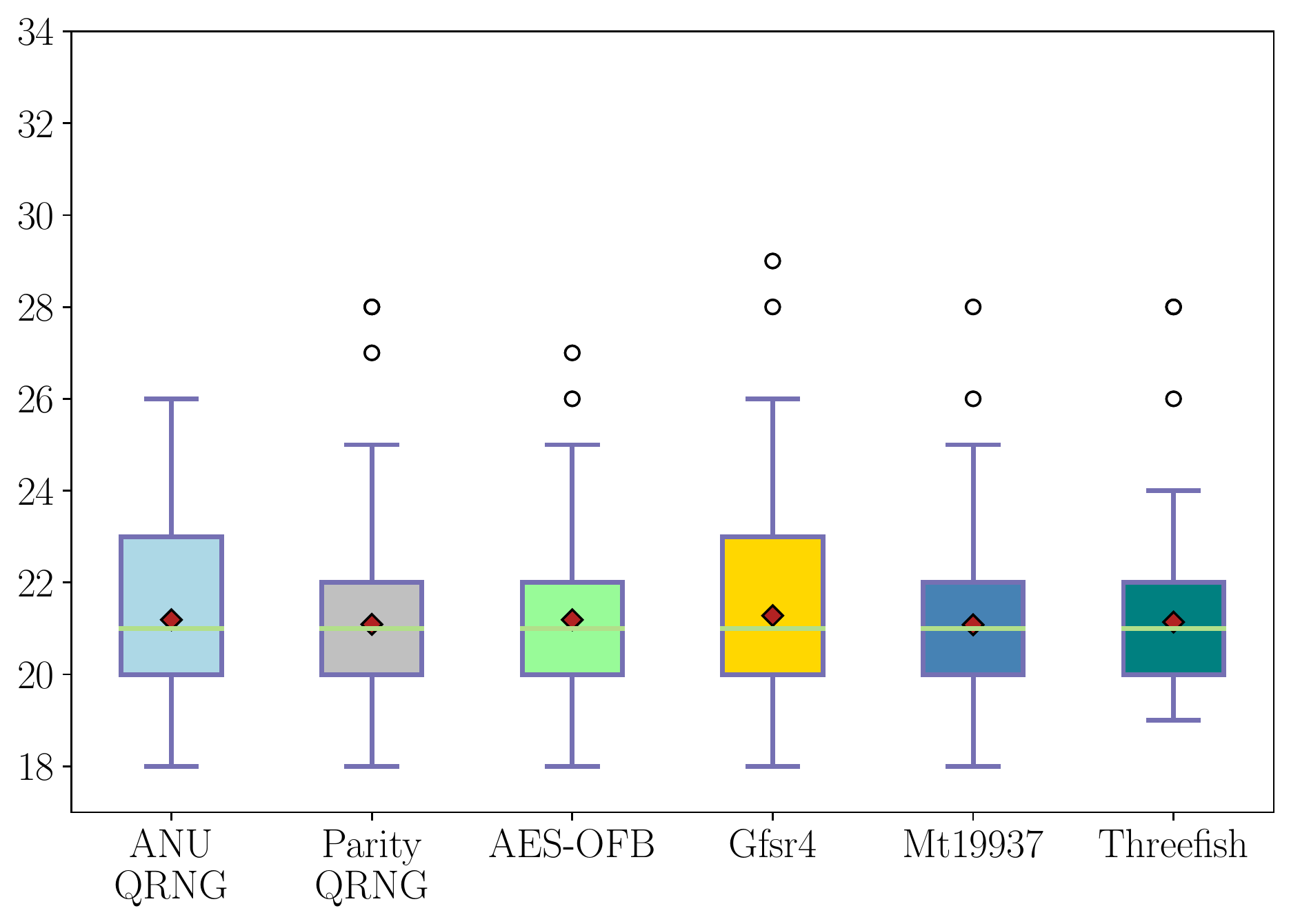} 
\caption{\label{fig:test1}Results of the first Chaitin--Schwartz--Solovay--Strassen test for the (a) original and (b) complemented bits. The data shows the minimum number of witnesses required to verify the compositeness of all 8,220,777 Carmichael numbers up to $10^{20}$.}
\end{figure} 
 
The results from the first CSSS test are shown in Fig.~\ref{fig:test1}. The Kolmogorov--Smirnov test did not find any statistically significant differences between the six datasets, neither for the original nor the complemented strings. The Shapiro--Wilk test indicated non-normality of all datasets, and therefore we did not apply Welch's $t$-test.

Despite being repetitive in their bit patterns, the strings from the Parity QRNG did not perform worse on this test than the strings from the other RNGs. In other words, the test appears to be insensitive to the repetitions. We offer one possible explanation. Reading sequentially through a looped string is akin to repeatedly reading through the non-looped (unique) portion of the string with a varying starting offset on each iteration. Therefore, even though the test repeatedly reads through the same 64,000-bit string, each iteration will typically produce a (mostly) fresh set of witnesses that cumulatively appear to be capable of delivering a performance for this test metric that is akin to that of the other RNGs.

\subsection{Second Chaitin--Schwartz--Solovay--Strassen test}

\begin{figure}
\centering

{\small (a)} 

\includegraphics[width=3.4in]{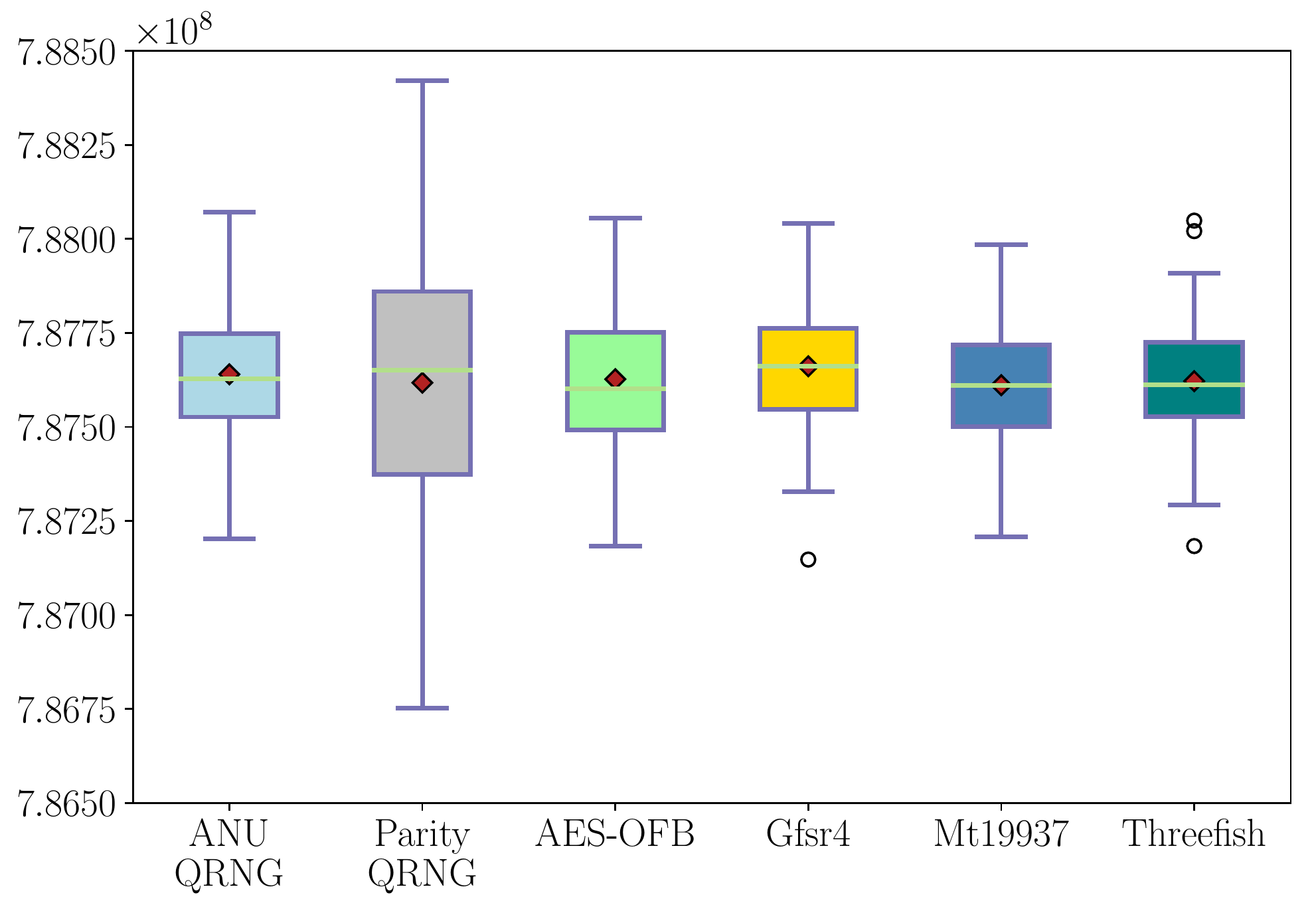} 

{\small (b)}

\includegraphics[width=3.4in]{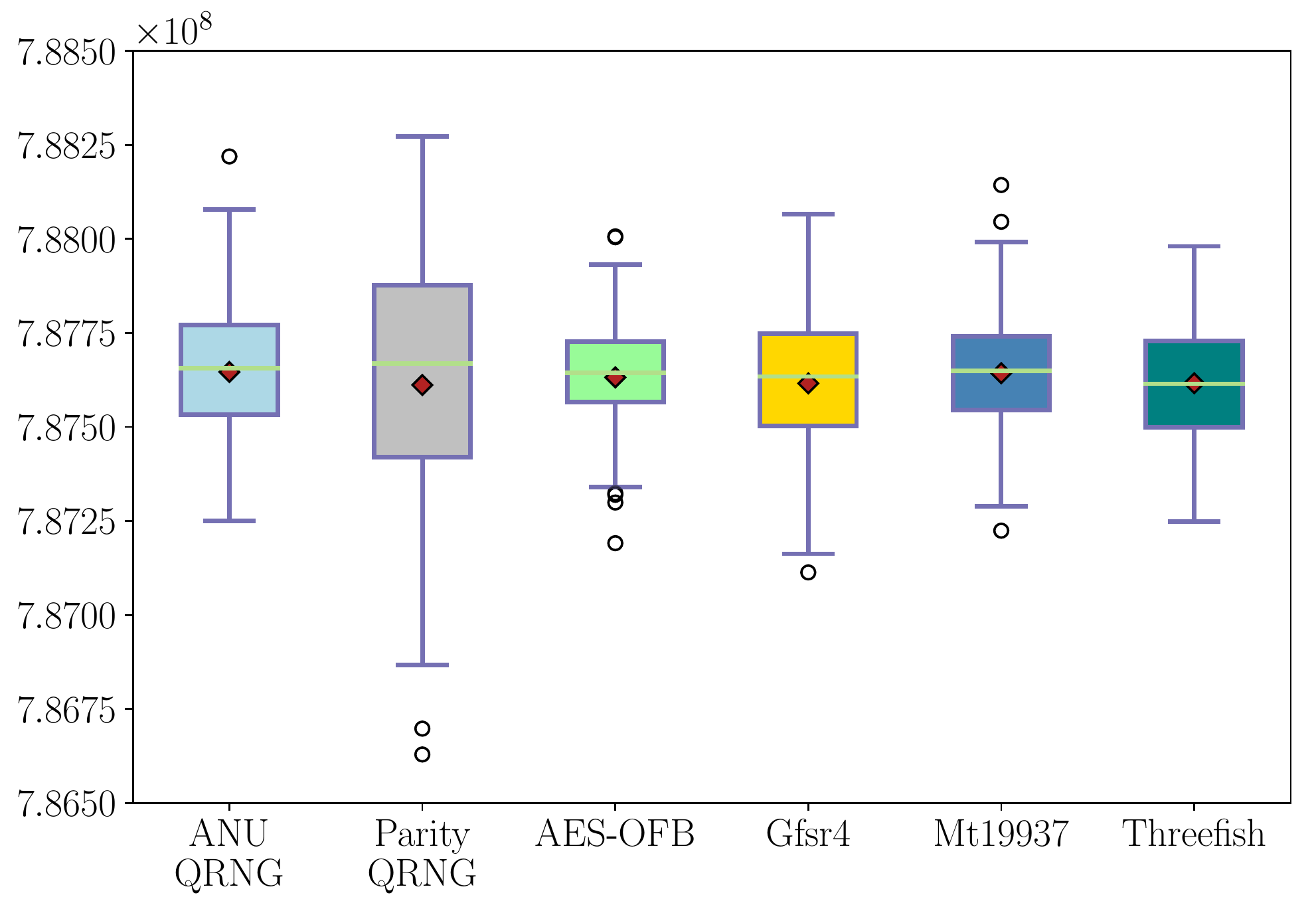} 
\caption{\label{fig:test2}Results of the second Chaitin--Schwartz--Solovay--Strassen test for the (a) original and (b) complemented bits. The data shows the number of bits required to successfully witness the compositeness of all 8,220,777 Carmichael numbers up to $10^{20}$.}
\end{figure} 
 
Figure~\ref{fig:test2} shows the results from the second CSSS test. The distribution for the Parity QRNG is seen to be broader than the other datasets. A possible explanation could be that because the unique portion of each Parity QRNG string is short and thus the set of distinct witnesses is comparably limited, variations in performance between sets of witnesses from different strings become more sharply accentuated.

While the Kolmogorov--Smirnov test identified statistically significant differences only between the Parity QRNG and the Gfsr4 PRNG ($p=0.0010$) for the original (noncomplemented) strings, we also found that $p$-values for pairs of datasets that involved the Parity QRNG were consistently much smaller than the $p$-values for other pairs of datasets, for both the original and the complemented strings. These observations may cautiously suggest that the test has some capability of identifying the lack of randomness of the Parity QRNG strings caused by their repeated structure. Of course, whether this finding points to an ability of the test to identify signatures of algorithmic randomness is questionable. Indeed, for the complemented strings, the Kolmogorov--Smirnov test no longer indicated statistically significant differences between the Parity QRNG and Gfsr4 datasets ($p=0.069$), even though, as mentioned above, interchanging 0s and 1s in the string should not affect algorithmic properties. 

The Shapiro--Wilk test confirmed normality of all datasets, and Welch's $t$-test failed to detect any statistically significant differences in the means between the datasets. 

\subsection{Third Chaitin--Schwartz--Solovay--Strassen test}

\begin{figure}
\centering

{\small (a)}

\includegraphics[width=3.4in]{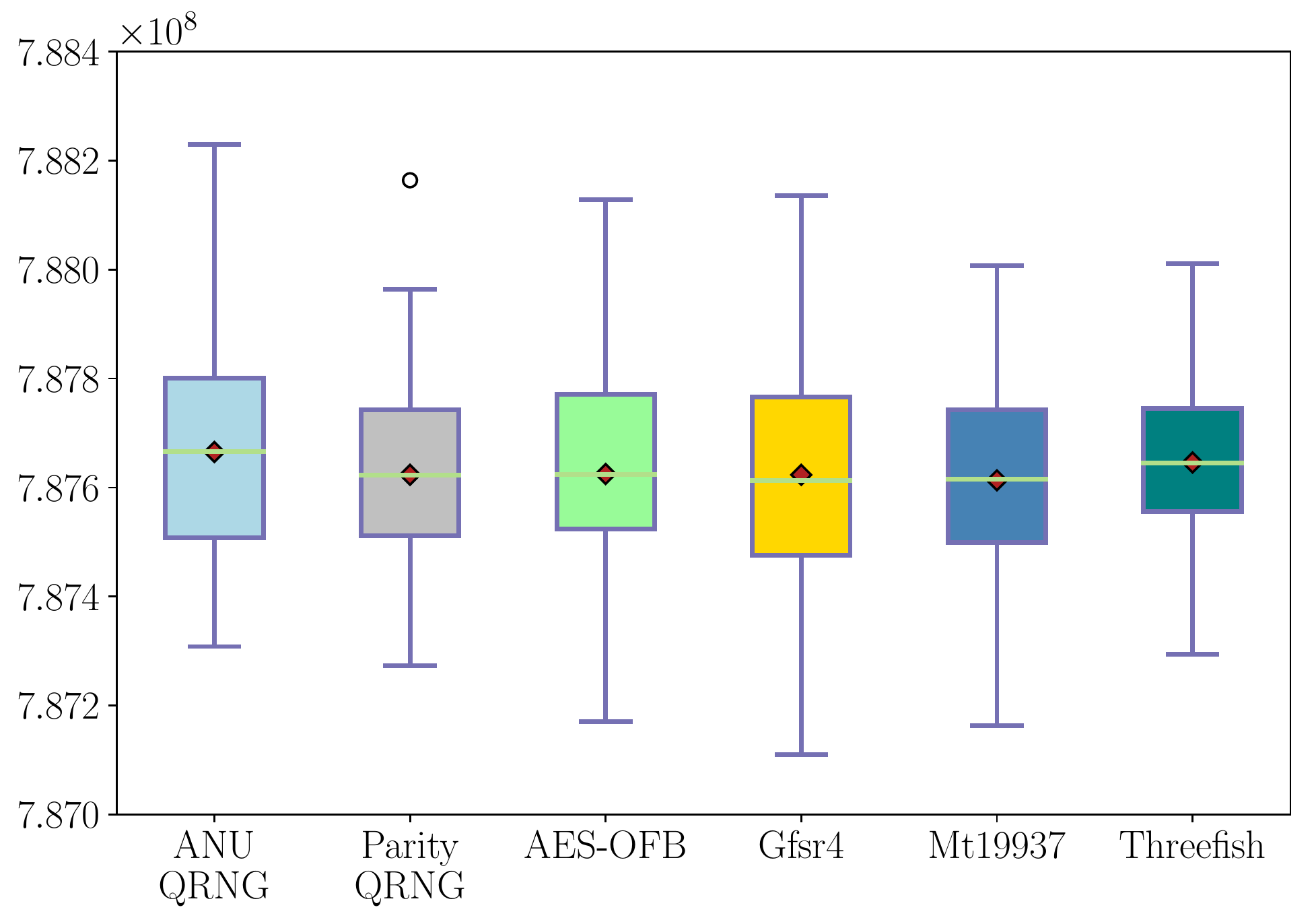} 

{\small (b)}

 \includegraphics[width=3.4in]{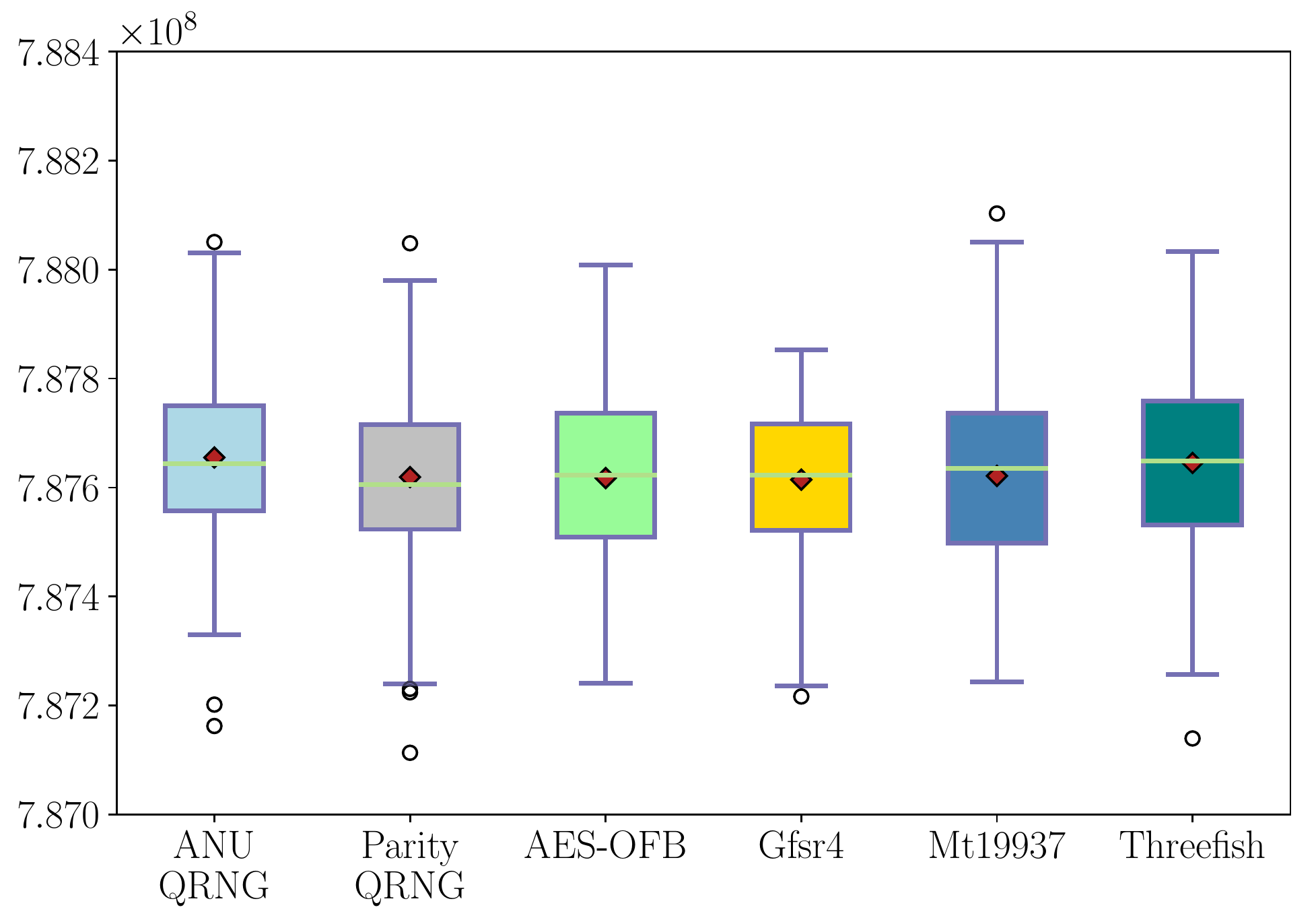} 
\caption{\label{fig:test3}Results of the third Chaitin--Schwartz--Solovay--Strassen test for the {\small (a)} original and {\small (b)} complemented bits. The data shows the number of bits required to successfully witness the compositeness of all 8,220,777 Carmichael numbers up to $10^{20}$, using the Chaitin--Schwartz compound predicate of Eq.~(\ref{eq:Z}).}
\end{figure} 

The results from the third CSSS test are shown in Fig.~\ref{fig:test3}.  No statistically significant differences were found by either the Kolmogorov--Smirnov test or Welch's $t$-test, whether for the original or the complemented strings. The Shapiro--Wilk test indicated normality of all datasets.

\subsection{\label{sec:fourth-chait-schw}Fourth Chaitin--Schwartz--Solovay--Strassen test}

\begin{figure}
\centering

{\small (a)} 

\includegraphics[width=3.4in]{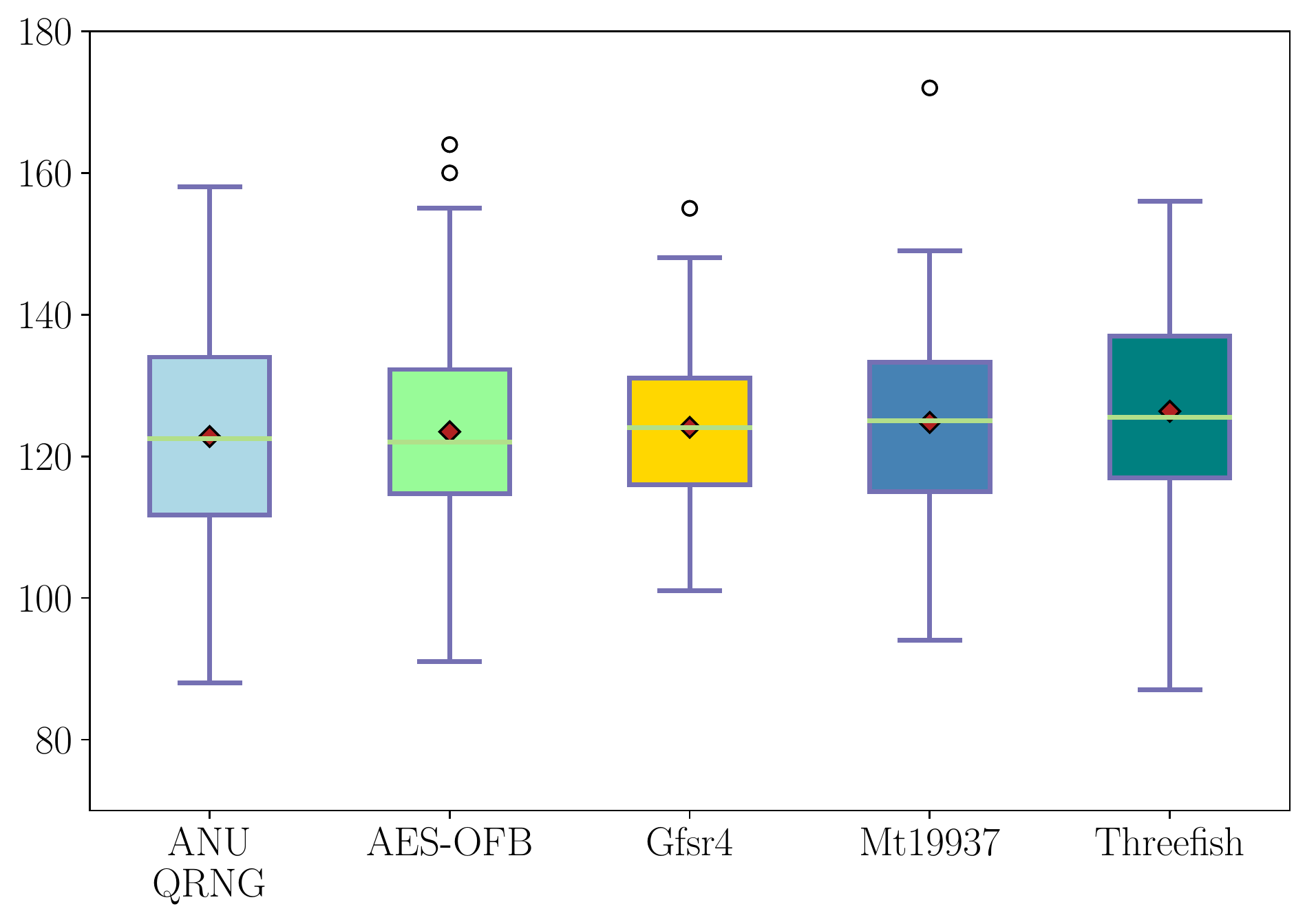} 

{\small (b)} 

\includegraphics[width=3.4in]{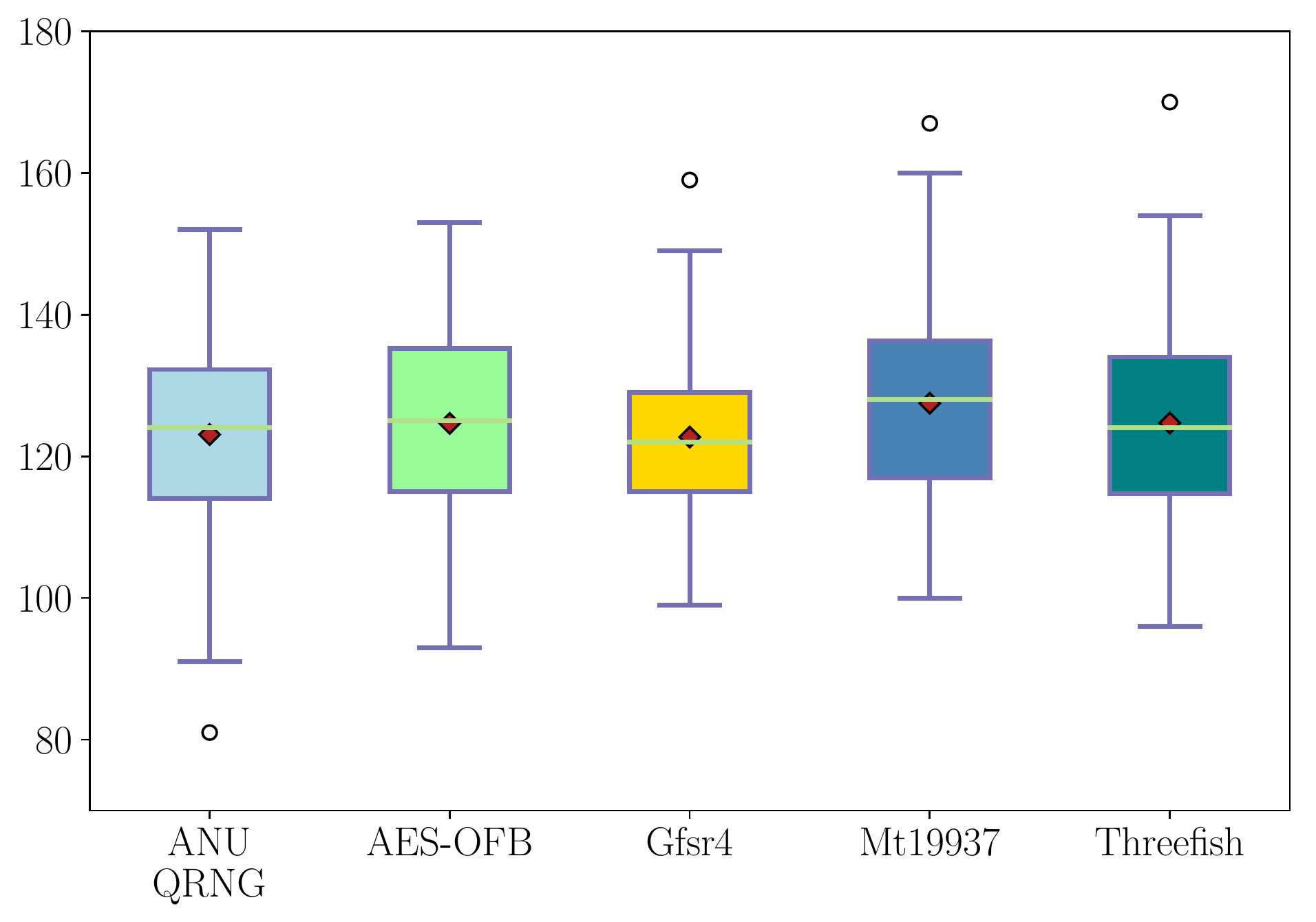} 
\caption{\label{fig:test4}Results of the fourth  Chaitin--Schwartz--Solovay--Strassen test for the (a) original and (b) complemented bits. The data shows the total number of observed violations of the Chaitin--Schwartz theorem for all 26 test numbers given by (\ref{eq:testnos}), accumulated over repeated passes through a given random string with an incremental starting offset. See main text for a description of the results for the Parity QRNG.}
\end{figure} 

The results from the fourth CSSS test are shown in Fig.~\ref{fig:test4}. For the noncomplemented strings from the Parity QRNG, none of the strings generated any violations. This can be understood by recalling that it is highly unlikely for a given witness string to produce a violation, and that therefore a large pool of unique witnesses is required to observe violations. The strings from the Parity QRNG, because of their looped content, did not provide such a sufficiently large pool. For the complemented strings from the Parity QRNG, one witness in one string produced a violation for the test number $n=9$. This gave rise to a total number of 26,214 violations for this particular string, since there are 26,214 complete repetitions of the unique (unlooped) portion (see Sec.~\ref{sec:parity-qrng}), with each such repeated portion producing the same violation-inducing witness. Given the isolated nature of this violation, it is difficult to provide a meaningful quantitative comparison of the results for the Parity QRNG and the other RNGs, and we have therefore not displayed these results in Fig.~\ref{fig:test4}. The results are nonetheless important, because they clearly show that the fourth CSSS test is by far the most sensitive of the CSSS tests when it comes to detecting the repetitive structure of the strings from the Parity QRNG. 

For the remainder of the datasets from the five other RNGs, no statistically significant differences were found by the Kolmogorov--Smirnov test, and the Shapiro--Wilk test indicated normality of the distributions. We note that the distributions of all test results are similar between the original and complemented bits. Thus we do not encounter the behavior observed in Ref.~\cite{Abbott:2019:uu}, where statistically significant differences between a QRNG and the PRNGs were found for the original but not for the complemented strings. 

\begin{table}
\caption{\label{tab:viol}Likelihood thresholds for the observation of violations of the Chaitin--Schwartz theorem. The table shows: each test number $n$; the corresponding length $m$ of the witness string in bits; the length $k$ of this string in base $(n-1)$; and the upper bound $p_{SS}=2^{-k}$ (not saturated) on the probability (per witness) of the occurrence of a violation as given by the Solovay--Strassen theorem. The last column shows the average relative frequency $p_\mathrm{obs}$ of violations observed when running the fourth CSSS test. The average is taken over the set of random strings, both original and complemented, produced by all RNGs except the Parity QRNG (we omit the Parity QRNG because its inability to produce violations is due to the use of looped strings).}
\begin{ruledtabular}
\begin{tabular}{lllll}
\vspace{.1cm}
$n$ & $m$ & $k$ & $p_{SS}$ &  $p_\mathrm{obs}$ \\
\hline\\[-.25cm]
 9 & 40 & 13 & $1.2 \times 10^{-4}$ & $6.0 \times 10^{-8}$ \\
15 & 40 & 10 & $9.8 \times 10^{-4}$ & $1.4 \times 10^{-8}$ \\
21 & 65 & 15 & $3.1 \times 10^{-5}$ & 0\\
25 & 65 & 14 & $6.1 \times 10^{-5}$ & $6.6 \times 10^{-11}$  \\
27 & 65 & 13 & $1.2 \times 10^{-4}$ & 0 \\
33 & 96 & 19 & $1.9 \times 10^{-6}$ & 0 \\
35 & 96 & 18 & $3.8 \times 10^{-6}$ & 0 \\
39 & 96 & 18 & $3.8 \times 10^{-6}$ & 0 \\
45 & 96 & 17 & $7.6 \times 10^{-6}$ & 0 \\
49 & 96 & 17 & $7.6 \times 10^{-6}$ & 0 \\
51 & 96 & 17 & $7.6 \times 10^{-6}$ & 0 \\
55 & 96 & 16 & $1.5 \times 10^{-5}$ & 0 \\
57 & 96 & 16 & $1.5 \times 10^{-5}$ & 0 \\
63 & 96 & 16 & $1.5 \times 10^{-5}$ & 0 \\
65 & 133 & 22 & $2.4 \times 10^{-7}$ & 0 \\
69 & 133 & 21 & $4.8 \times 10^{-7}$ & 0 \\
75 & 133 & 21 & $4.8 \times 10^{-7}$ & 0 \\
77 & 133 & 21 & $4.8 \times 10^{-7}$ & 0 \\
81 & 133 & 21 & $4.8 \times 10^{-7}$ & 0 \\
85 & 133 & 20 & $9.5 \times 10^{-7}$ & 0 \\
87 & 133 & 20 & $9.5 \times 10^{-7}$ & 0 \\
91 & 133 & 20 & $9.5 \times 10^{-7}$ & 0 \\
93 & 133 & 20 & $9.5 \times 10^{-7}$ & 0 \\
95 & 133 & 20 & $9.5 \times 10^{-7}$ & 0 \\
99 & 133 & 20 & $9.5 \times 10^{-7}$ & 0 \\
561 & 280 & 30 & $9.3 \times 10^{-10}$ & 0 
\end{tabular}
\end{ruledtabular}
\end{table}

We observed that the majority of violations were produced for the smallest two numbers tested, $n=9$ and $n=15$. Several strings also produced violations for $n=25$, but no more than one violation per string. None of the other numbers $n$ we have tested [see (\ref{eq:testnos})] produced any violations. To analyze these observations, recall from Sec.~\ref{sec:chait-schw-solov} that the Solovay--Strassen theorem implies that the probability for a witness string $s$ with $k$ digits in base $(n-1)$ to produce a violation is bounded from above by $2^{-k}$. Table~\ref{tab:viol} shows the values of these Solovay--Strassen bounds for all test numbers $n$. We can compare these bounds to the observed relative frequencies $p_\mathrm{obs}$  of violations for the strings produced by the different RNGs. Each string has $N$ bits, for each test of a violation an $m$-bit witness is read, and a total of $m$ passes (with incremental offset) are made through the random string. Thus, there are a total of $N$ witnesses for which a violation is checked. This implies $p_\mathrm{obs}=N_\mathrm{viol}/N$ for a given random string, where $N_\mathrm{viol}$ is the observed number of violations for $n$ over all $m$ passes through the string. 

The data in Table~\ref{tab:viol} clearly show that the observed average frequencies of violations fall below the Solovay--Strassen bounds by several orders of magnitude; indeed, as mentioned, for most $n$ those frequencies are zero as no violations were obtained. If the values of the Solovay--Strassen bounds were a good indicator of the likelihood of actually finding violations, then, given our string lengths $N$, we would expect to see violations for all $n$. For example, for the largest number, $n=561$, the Solovay--Strassen bound is $2^{-30} \approx 9.3 \times 10^{-10}$, which translates into a minimum of 180 passes through the $N$-bit random string (on average) to produce a violation. We made 280 passes but observed no violations for any of the 600 strings we have tested. Indeed, the Solovay--Strassen bound appears to be, at best, an incomplete indicator of the likelihood of observing a violation. For example, $n=9$ and $n=27$ have the same Solovay--Strassen bound, but $n=27$ produced no violations while $n=9$ produced many. Similarly, $n=27$ has a larger Solovay--Strassen bound than $n=25$, and yet $n=25$ produced occasional violations while $n=21$ produced none. 

\section{\label{sec:discussion}Discussion}

We considered five tests of randomness: a test of Borel normality, and four versions of the Chaitin--Schwartz--Solovay--Strassen tests (developed in Ref.~\cite{Abbott:2019:uu}) that probe algorithmic randomness and incomputability. We applied these tests to samples of 100 long ($25 \times 2^{26}$ bits) strings produced by two QRNGs and four PRNGs. The tests did not find any statistically significant differences in performance on the tests between the QRNGs and PRNGs that would point to evidence of algorithmic randomness and incomputability in the output from the QRNGs. While several of the test results did differ for the strings from the Parity QRNG, these differences could be attributed to the much shorter lengths of these strings, which necessitated extensive recycling of bits.

Our results may help both confirm and clarify the results of Ref.~\cite{Abbott:2019:uu}. There, the tests mostly failed to find statistical differences between a QRNG and a set of five PRNGs, with two exceptions. The QRNG behaved differently on the second CSSS test, though the difference could be shown (by demonstrating reversed behavior for the complemented strings) to be caused by bias in the output from the QRNG. The QRNG also behaved differently on the fourth CSSS test. It did so, however, only for the original (noncomplemented) strings, and the interpretation of the results remained inconclusive. In our study, neither of the QRNGs exhibited significant bias, owing to postprocessing of the output from the ANU QRNG, and owing to the use of the parity variable for the bit generation in the case of the Parity QRNG. Thus, we expect that bias is not a significant factor in our tests. This is confirmed by the observation that the test results for the original and complemented strings were very similar, and that, except in one isolated case, there were no statistically significant differences between RNGs that would appear for  either only the original or only the complemented strings.

Despite being composed of thousands of repetitions of comparably short unique strings, the strings from the Parity QRNG did not exhibit statistically significant differences in performance on the first and third CSSS tests. The second test, on the other hand, seemed to show some sensitivity to the repetitions in the strings, producing $p$-values for pairs of generators that were consistently lower when the Parity QRNG was involved in the comparison. The most striking discrepancy was observed for the fourth test, for which the entire collection of unique bits from the Parity QRNG  generated only a single violation. This can be explained by noting the very small likelihood that a violation occurs for a given witness, combined with the comparably small set of distinct witnesses obtained from the Parity QRNG. 

Of all the four CSSS tests, the fourth test appears to be the most promising, simply because the Chaitin--Schwartz theorem on which it is based directly appeals to the property of $c$-Kolmogorov randomness. Its ability to detect the repeated structure of the strings from the Parity QRNG may also be regarded as an advantage. As was already clearly recognized in Ref.~\cite{Abbott:2019:uu}, the main drawback of the test is that the violations of the Chaitin--Schwartz theorem searched for by the test happen very rarely, and thus one needs to have available very long strings. The vast majority of the violations we were able to observe were for the smallest numbers we tested ($n=9$ and $n=15$). If one aims to observe violations for even the smallest Carmichael numbers, it appears that much longer strings than those we have used would be required, as well as an amount of computing time that might quickly become prohibitively large. It would be desirable to develop a test of algorithmic randomness that, like the fourth CSSS test, appeals to the Chaitin--Schwartz theorem, but implements it in a manner that does not require exorbitant resources in terms of string lengths and time.

\subsection*{Data availability}

We have made available all data used in this paper (random strings, computer code, and test results) at \url{http://faculty.up.edu/schlosshauer/randomness}.

\begin{acknowledgments} 

We thank R.\,G.\,E.\ Pinch for providing us with the set of Carmichael numbers \cite{Pinch:2007:km}. We are grateful to the authors of Ref.~\cite{Abbott:2019:uu} for making publicly available their test code, and to the Centre for Quantum Computation and Communication Technology at the Australian National University for making publicly available their random strings. This work was supported by the M.\,J. Murdock Charitable Trust  (NS-2015298) and the SURE program of the University of Portland.

\end{acknowledgments} 


%

\end{document}